\documentclass[12pt,epsf,epsfig,psfig]{article}
\usepackage{graphicx}
\usepackage{epsfig}
\def\J{$J/\psi$}
\def\j{J/\psi}
\def\X{$\chi_c$}

\def\P{$\psi'$}

\def\be{\begin{equation}}
\def\ee{\end{equation}}

\def\lsim{\raise0.3ex\hbox{$<$\kern-0.75em\raise-1.1ex\hbox{$\sim$}}}
\def\gsim{\raise0.3ex\hbox{$>$\kern-0.75em\raise-1.1ex\hbox{$\sim$}}}

\def\NP{{ Nucl.\ Phys.\ }}
\def\PL{{ Phys.\ Lett.\ }}

\def\PRL{{ Phys.\ Rev.\ Lett.\ }}

\def\ZP{{ Z.\ Phys.\ }}

\begin{document}

\hfill BI-TP 2003/30 

\thispagestyle{empty}

\vskip 1cm

\centerline{\Large \bf Predictions for \J~Suppression}

\bigskip

\centerline{\Large \bf by Parton Percolation}

\vskip 1cm 

\centerline{\large
S.\ Digal$^a$, S.\ Fortunato$^a$ and H.\ Satz$^{a,b}$} 

\vskip 1cm

\medskip

\centerline{a: Fakult\"at f\"ur Physik, Universit\"at Bielefeld}

\centerline{Postfach 100 131, D-33501 Bielefeld, Germany}
 
\medskip

\centerline{b: Centro de Fis{\' \i}ca das Interac{\c c}{\~o}es 
Fundamentais (CFIF)}

\centerline{Instituto Superior T\'ecnico, Av. Rovisco Pais, P-1049-001 Lisboa, 
Portugal}

\vskip 1cm

\centerline{\large \bf Abstract:}

\bigskip

Parton percolation provides geometric deconfinement in the
pre-equilibrium stage of nuclear collisions. The resulting parton
condensate can lead to charmonium suppression. We formulate a
local percolation condition viable for non-uniform collision 
environments and show that it correctly reproduces the suppression observed 
for $S\!-\!U$ and $Pb\!-\!Pb$ collisions at the SPS. Using this 
formulation, we then determine the behavior of \J~suppression 
for $In\!-\!In$ collisions at the SPS and for $Au\!-\!Au$ collisions 
at RHIC.

\newpage

\noindent{\bf \large 1.\ Introduction}

\bigskip

In recent years, the role of the partonic initial stages of high energy 
nuclear collisions has attracted increasing interest. While the ultimate
aim of the experimental program is the production of the quark-gluon
plasma predicted by statistical QCD, it is becoming more and more
evident that a thermalized medium consisting of quarks and gluons can 
be produced only if already the initial state of the collision provides 
the conditions necessary for deconfinement and thermalization. Dilute
initial state parton configurations do not lead to the colour connection
needed so that partons from different collisions can combine to
form a collective medium. On the other hand, if the initial state of 
primary partons is sufficiently dense, 
cluster percolation will set in, leading to a condensate of connected 
and hence interacting partons, which are no longer associated to any 
``parent'' hadrons. Thus parton percolation is a geometric, pre-equilibrium 
form of deconfinement \cite{perco,torino}; it must occur in any description
based on partons of an intrinsic transverse momentum.  

\medskip

Subsequent interactions can thermalize this interconnected partonic system, 
turning it into a quark-gluon plasma, so that parton percolation constitutes 
an essential pre\-re\-qui\-site for QGP formation. Moreover, pre-equilibrium 
deconfinement leads to a further important  
question. Much of the investigation of high energy nuclear collisions is 
devoted to the search for quark-gluon plasma signatures. However, some of 
the features observed in nuclear interactions might be determined by the 
system present before any equilibration has occurred, and they could be 
independent of a subsequent thermalization. We must therefore investigate
observable consequences of parton percolation, and if possible,
show how they can be distinguished from those due to QGP formation.

\medskip

One of the features of particular interest in this connection is the 
suppression of \J~production in nuclear collisions, predicted as
deconfinement signal for thermal media quite some time ago \cite{MS}. If 
the parton condensate formed through percolation contains partons hard 
enough to resolve the produced charmonium states, these partons can also 
dissociate charmonia, so that the onset of \J~suppression could coincide 
with that of parton percolation. It was shown in a first study \cite{DFPS}
that for $Pb\!-\!Pb$ collisions at the CERN-SPS as function of centrality 
the resulting thresholds appear quite reasonable. A more quantitative
comparison, which can then also provide the basis for further predictions,
requires a formulation of percolation in a non-uniform environment, as 
given by the partonic source profile in nuclear collisions, and adapted
to the local nature of charmonium probes. The aim of this paper is first 
to formulate percolation for such conditions, then to compare the results
to existing SPS data ($S-U$ and $Pb-Pb$), and finally to present the 
predictions of this approach for the forthcoming \J~production experiments 
at CERN and BNL. All results are obtained by Monte Carlo simulations
taking into account the centrality dependence of the collisions.
In closing, we shall consider some features which could
distinguish \J~suppression by parton percolation from a suppression in
a thermal medium.
 
\bigskip

\noindent{\bf \large 2.\ Local Parton Percolation Conditions}

\bigskip

\begin{figure}[htb]
\mbox{
\epsfig{file=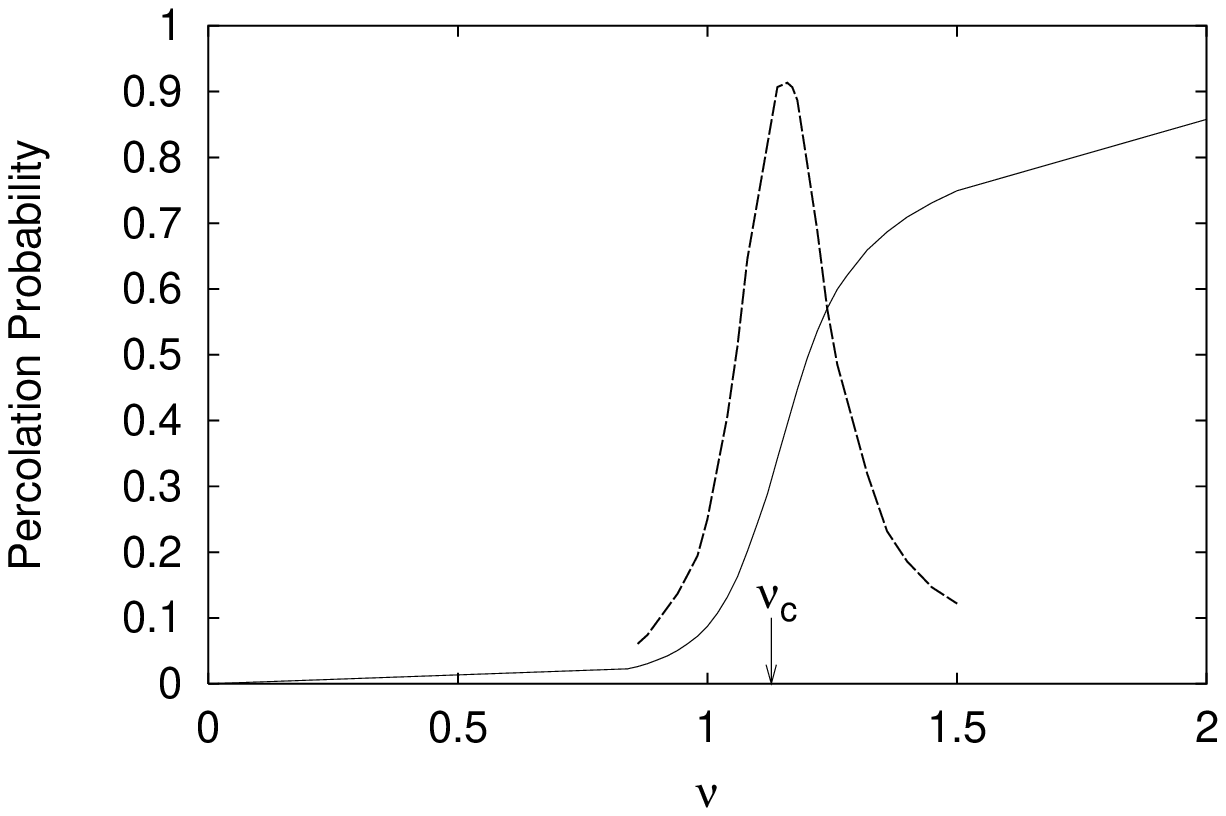,width=7.3cm}
\hskip0.9cm
\epsfig{file=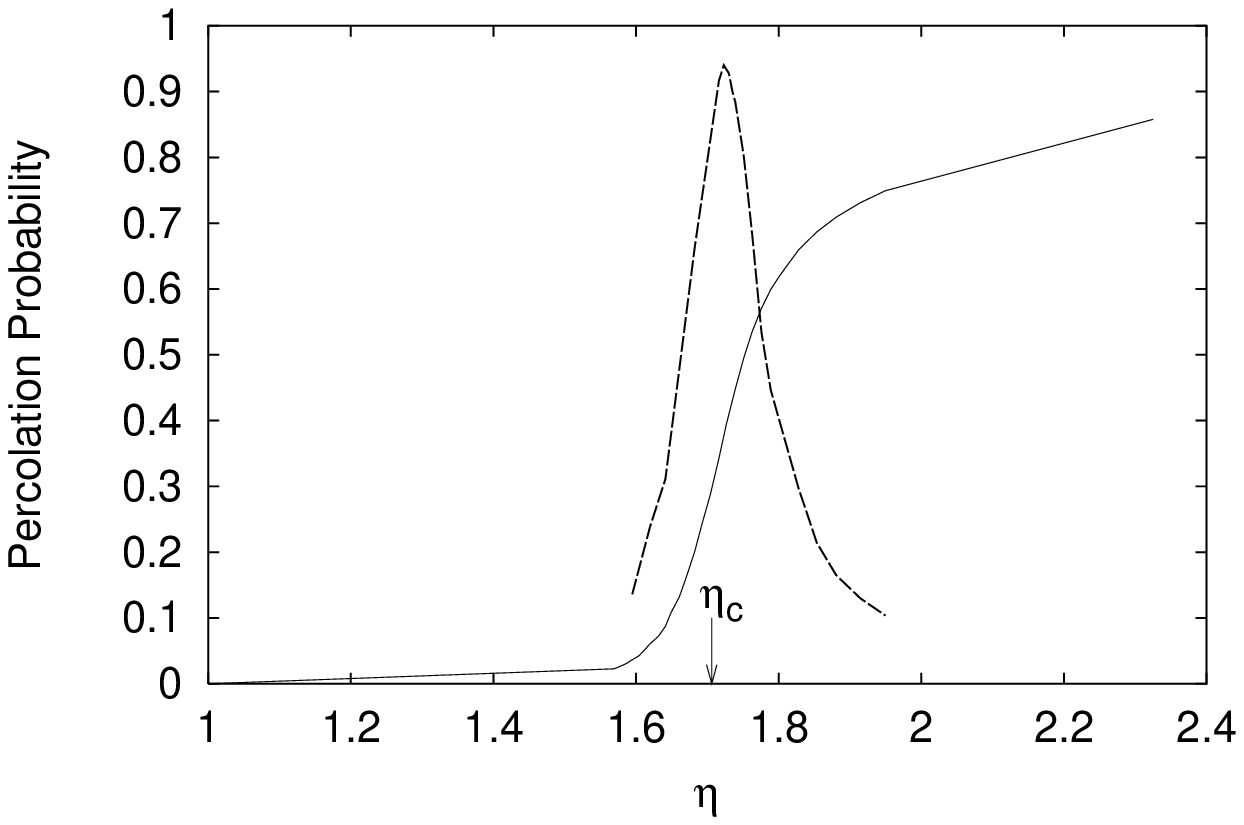,width=7.3cm}}
\vskip -4.5cm
{\hspace*{1.6cm} (a) \hskip 7.7cm (b)} 
\vskip 3.7cm
\caption{Cluster growth and corresponding derivative as function of the 
overall filling factor $\nu$ (a) and of the cluster filling factor $\eta$ 
(b).}
\label{perco}
\end{figure}

Consider a flat two-dimensional circular surface of radius $R$ (the 
transverse nuclear area), on which $N$ small discs of radius $r\ll R$ 
(the transverse partonic size) are randomly distributed, allowing overlap.
With increasing density $n\equiv N/\pi R^2$, clusters of increasing size
appear. The crucial feature is that this cluster formation shows
critical behaviour: in the limit $N \to \infty$ and $R \to \infty$, 
with $n$ finite, the cluster size diverges at a certain critical density 
$n=n_c$. This percolation threshold is given by
\be
n_c = {\nu_c \over \pi r^2};
\label{1}
\ee
the critical value $\nu_c \simeq 1.13$ of the `filling factor'
$\nu \equiv n(r/R)^2$ is determined by numerical studies. For finite $N$ 
and $R$, percolation sets in when the largest cluster spans the entire 
circular surface from the center to the edge. The resulting cluster growth 
is illustrated in Fig.\ \ref{perco}a for a ratio $r/R=1/100$; we show the 
{\sl percolation probability}, defined as the relative size of the largest
cluster, as function of the filling factor $\nu$. Because of overlap, 
a considerable fraction of the surface is still empty at the percolation 
point; in fact, at that threshold, only $1-\exp\{-\nu_c\} \simeq 2/3$ of 
the surface is covered by discs.

\bigskip

\medskip

In high energy nuclear collisions, the incident partons are distributed
on a plane transverse to the collision axis. However, since they originate
from the nucleons within the colliding nuclei, this distribution is highly 
non-uniform, with more nucleons and hence more partons in the center 
than towards the edge of the transverse nuclear plane. If the basic surface 
is not flat, it is still possible to define the percolation threshold as 
the point at which the cluster size shows singular behavior in the limit of 
large $R$ and $N$ \cite{torino}. Such a {\sl global} definition is, 
however, not always the most useful. Hard probes, such as quarkonia,
probe the medium locally and thus test only if it has reached the
percolation point and the resulting geometric deconfinement at their
location; they cannot register the global features of the
entire cluster area. It is thus necessary to define a more {\sl local} 
percolation criterium, and this is in fact quite straight-forward.  

\medskip

As mentioned, at the percolation point, $\exp\{-\nu_c\} \simeq 1/3$
of the surface remains empty. Hence the disc density in the percolating 
cluster must be greater than $(1.5~ \nu_c/\pi r^2)$. Numerical studies 
show that in fact percolation sets in when the density $m$ of constituents 
in the largest cluster reaches the critical value 
\be
m_c = {\eta_c \over \pi r^2},
\label{2}
\ee  
with $\eta_c \simeq 1.72$, slightly larger than $1.5~ \nu_c$. This 
result provides the required local test: if the parton density at a certain 
point in the transverse nuclear collision plane has reached this level, 
the medium there belongs to a percolating cluster and hence to a deconfined 
parton condensate. In Fig.\ \ref{perco}b, we illustrate how the percolating 
cluster grows as function of the cluster filling factor 
$\eta=m(r/R)^2$, again with $r/R=1/100$.

\medskip

Let us recapitulate: there are two equivalent criteria which can be used
to specify the onset of percolation. In the `global' definition, it 
occurs when the total number of partons distributed over the entire 
transverse nuclear area reaches $1.13 / \pi r^2$. In the `local' 
definition, percolation sets 
in when the parton density in the largest cluster reaches $1.72/ \pi r^2$. 
Which of the two is preferable in a given case depends on the physics 
question addressed, and if we want to study local phenomena such as \J 
~suppression, it is clearly the local parton density which is relevant, 
not the average over the entire transverse area.

\medskip

We now turn to the implications of the approach to nuclear collisions
\cite{DFPS}; for illustration, we concentrate for the moment on central 
$A\!-\!A$ interactions at a c.m.s.\ energy $\sqrt s$ per nucleon-nucleon 
collision. The distribution of nucleons in the colliding nuclei is 
speci\-fied 
by a Glauber calculation using Woods-Saxon nuclear distributions \cite{KLNS};
this provides the density $n_s(A)$ of nucleons in the transverse collision 
plane. The parton content of a nucleon 
is given by parton distribution functions $dN_q(x,Q^2)/dy$ determined 
in deep inelastic scattering experiments; here $x$ denotes the fraction 
of the nucleon momentum carried by the parton and $Q$ the momentum 
resolution scale. At central c.m.s.\ rapi\-di\-ty $y =0$,
we have $x=k_T/\sqrt s$,
where $k_T$ denotes the average transverse momentum of the parton.
In nuclear collisions, $k_T$ defines the transverse size of the partons 
and thus also sets the resolution scale, $k_T \simeq Q$. Using these
quantities, we have for the `global' percolation condition in nuclear
collisions
\be
n_s(A) \left( {dN_q(x,Q_c^2) \over dy} \right)_{x=Q_c/\sqrt s}
= {\nu_c \over (\pi/Q_c^2)};
\label{3}
\ee
it determines for what value of $A$ at a given $\sqrt s$ percolation sets
in, and it specifies the value $Q=Q_c(A,\sqrt s)$ at the onset point.
Using the nuclear source density $n_s(A)$ together with the parton
distribution function, we can also calculate the density of partons 
$m(A,Q,\sqrt s)$ for the largest cluster in the transverse
collision plane. The relation
\be 
m_c(A,Q_c,\sqrt s) = {\eta_c \over (\pi/Q_c^2)};
\label{4}
\ee
then provides the `local' parton percolation condition to specify
the onset values of $A,\sqrt s$ and $Q$. These considerations 
can be extended to non-central collisions in a straight-forward
fashion \cite{DFPS,KLNS}.

\medskip

It is not {\sl a priori} evident that the threshold values $\nu_c$ and
$\eta_c$ for the non-uniform distribution specified by the Woods-Saxon 
distributions remain the same as for a uniform distribution, nor
is it clear how large finite size corrections are at given values of 
$r/R$. We have therefore calculated the pseudo-critical values
(derivative peak positions) of $\nu$ and $\eta$ for $R/r=20,~50$ and 
$100$, using the Woods-Saxon distribution for central $Pb-Pb$ collisions 
to determine the parton density as well as the average overall 
interaction region. Finite-size scaling techniques from statistical
physics then provide the asymptotic limit; in this way, we have verified 
that $\nu_c$ and $\eta_c$ for this non-uniform distribution in fact agree
with the above mentioned values for the uniform case. Moreover, we
found that for the mentioned values of $R/r$, the deviations from the
asymptotic critical values are less than 2 \%.   

\bigskip

\noindent{\bf \large 2.\ Comparison to Pb-Pb and S-U Data}

\bigskip 

We now want to study parton percolation for CERN-SPS conditions and 
compare the resulting consequences for \J~suppression to the data 
of the NA38/NA50 collaborations.

\medskip

\begin{figure}[htb]
\mbox{
\epsfig{file=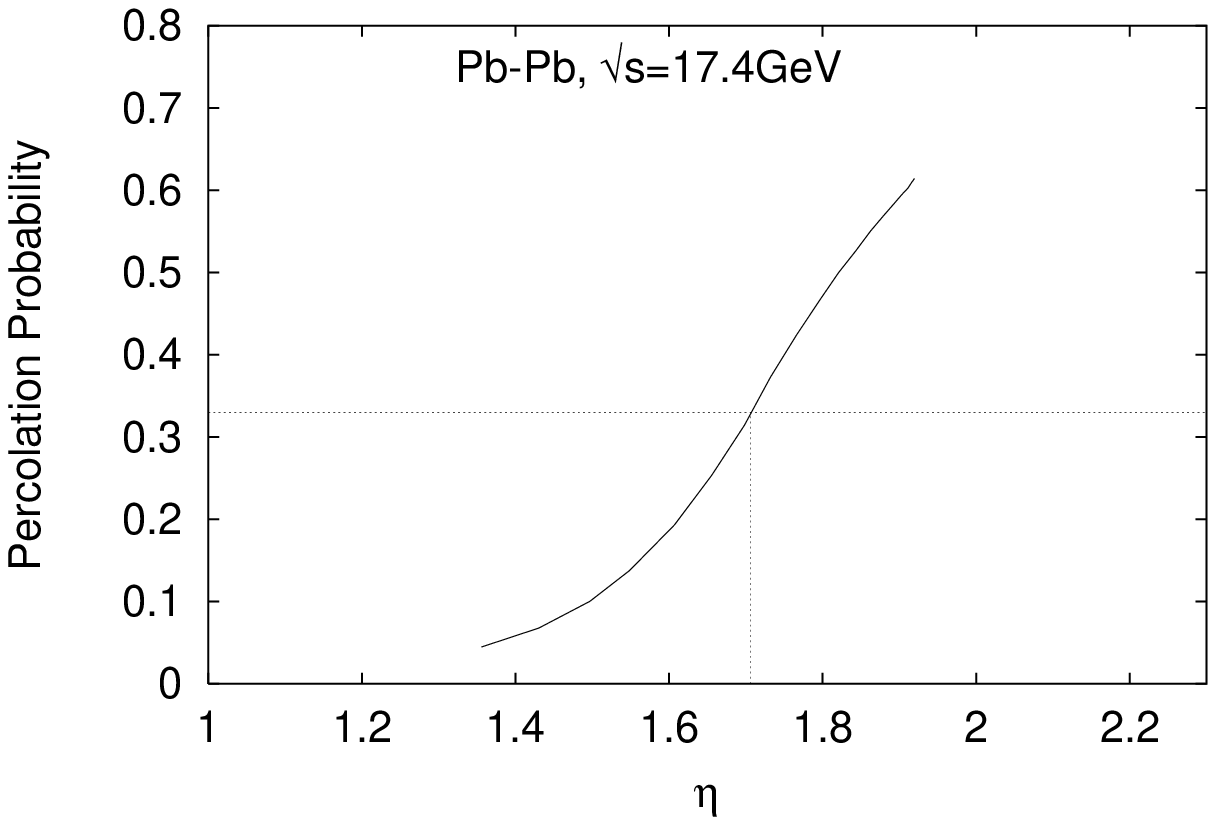,width=7.5cm}
\hskip0.5cm
\epsfig{file=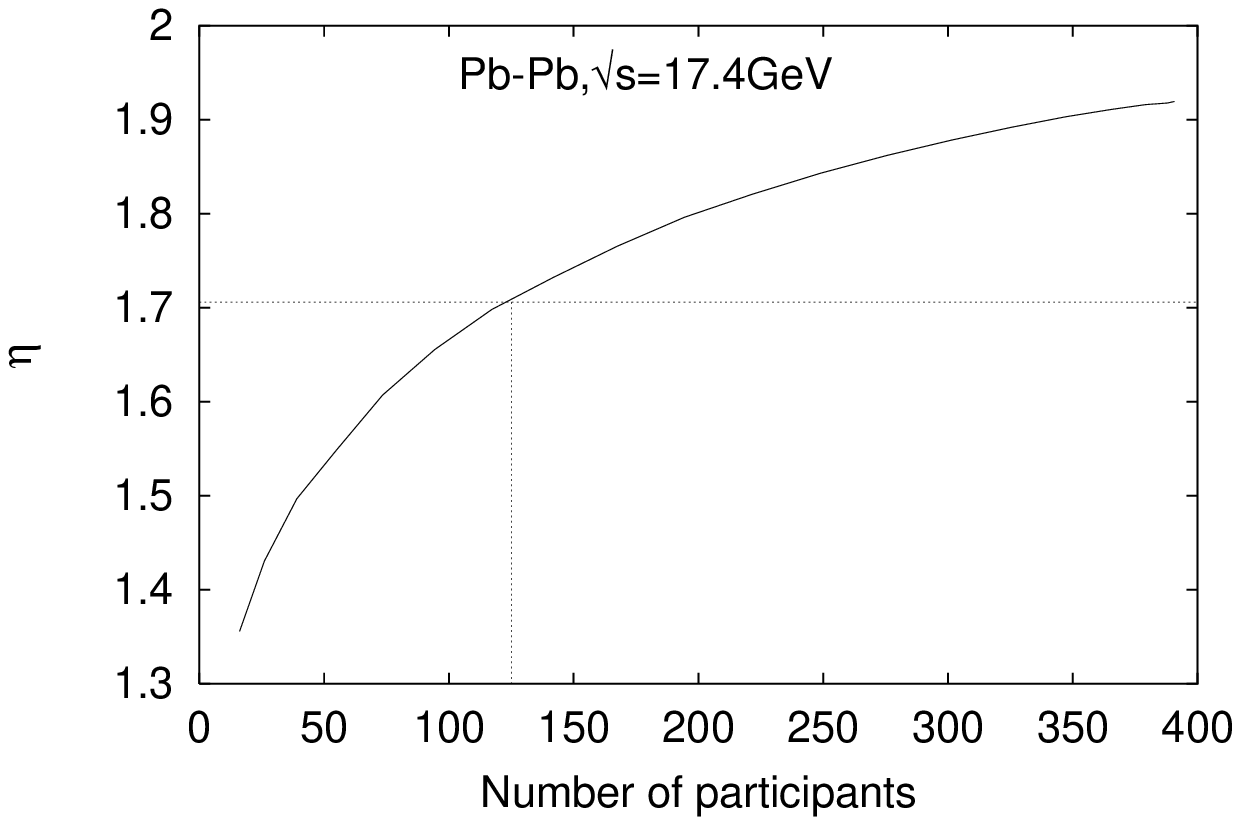,width=7.5cm}}
\vskip -4.6cm
{\hspace*{1.6cm} (a) \hskip 7.7cm (b)} 
\vskip 3.7cm
\caption{Percolation onset in Pb-Pb collisions at the SPS.}
\label{Pb-chi}
\end{figure}

The parton content of a nucleon is obtained from GRV94 parton 
distribution functions \cite{GRV94}, since these are available
for the required kinematic range; for more details, see \cite{DFPS}. 
With this specified, we calculate the size of the largest cluster 
as well as the overall size of the interaction region for different 
centralities (the latter is here defined by the ratio of the
number to the density of wounded nucleons).
For $Pb\!-\!Pb$ collisions at $\sqrt s = 17.4$ GeV, we show 
in Fig.\ \ref{Pb-chi}a the relative size of the largest cluster as function
of the filling factor $\eta$. It is seen that the cluster size grows 
rapidly in a narrow band around $\eta_c$, and at the percolation onset, 
the percolating cluster covers about 1/3 of the total collision area. 
In Fig.\ \ref{Pb-chi}b, we show the cluster filling factor $\eta$
as function of centrality, measured by the number of participating 
nucleons. The critical value for parton percolation, $\eta_c=1.72$, 
is found to be attained for $N_{\rm part}\simeq 125$.

\medskip

\medskip

The parton distribution functions specify both the number of 
partons and the resolution scale $Q_c$ at the percolation point.
For $Pb\!-\!Pb$ collisions at $\sqrt s = 17.4$ GeV, $Q_c \simeq
0.7$ GeV. The scales of the charmonium states \X~and \P~, as
determined by the inverse of their radii calculated in potential theory, 
are around 0.6 GeV and 0.5 GeV, 
respectively. Since the parton condensate can thus resolve these 
states, we assume, following \cite{DFPS}, that it will dissociate
them. To be specific: we assume that all \X~and \P~states formed
inside the percolating cluster disappear; the location of their 
formation is determined by the collision density as given by the 
Glauber formulation \cite{DFPS,KLNS}.
The first onset of \J~suppression in $Pb\!-\!Pb$ collisions 
at the SPS should therefore occur at $N_{\rm part}\simeq 125$, where
the \J's due to feed-down from \X~and \P~states in the percolating 
cluster are eliminated. Directly produced \J's survive because of their
smaller radii (leading to a scale of 0.9 - 1.0~ GeV), as do those
coming from excited states outside the percolating cluster. We
emphasize that with parton distribution functions and charmonium radii 
given, the description contains no adjustable parameters. 

\medskip

\begin{figure}[htb]
\hskip -0.5cm
\centerline{\psfig{file=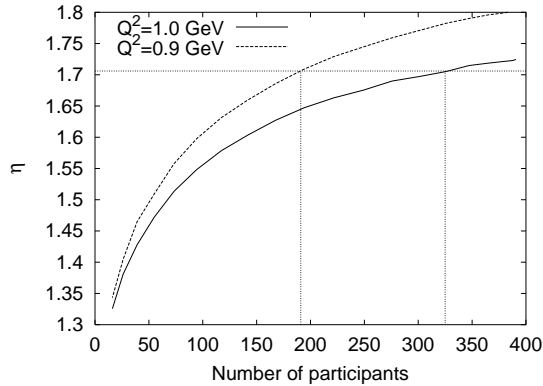,width=7.5cm}}
\caption{Percolation onset for directly produced \J's
in Pb-Pb collisions at the SPS.}
\label{dir-psi}
\end{figure}

As noted, the parton condensate formed at the onset of percolation 
has a resolution scale $Q_c \simeq 0.7$ GeV and thus cannot resolve
the ground state charmonium \J~with its smaller radius. The dissociation
of directly produced \J's thus requires more central collisions, which
lead to a higher parton density and simultaneously to a better
resolution, i.e., to an increase of $Q_c$. The precise onset point
is here more difficult to determine, since it is fixed by the value of
the \J~radius. For $r_{\j} = 0.20$ fm, we have $Q\simeq 1.0$ GeV and obtain
the centrality dependence shown in Fig.\ \ref{dir-psi}, indicating
that the suppression of directly produced \J's starts at 
$N_{\rm part} \simeq 320$. However, here a note of warning has to be 
added. The \X~and \P~states can be resolved by the parton
condensate when it is first formed; the values of their intrinsic
scales do not enter explicitly, since they are below the resolution scale
$Q_c$ of the condensate. The onset point for the suppression of
the $1S$ ground state is directly determined by the radius of that
state, and since the dependence of parton density on centrality
(see Fig.\ \ref{dir-psi}) is quite flat, small changes in  $r_{\j}$
lead to large changes in the threshold value. For instance, $r_{\j} = 
0.22$ fm ($Q\simeq 0.9$ GeV)
shifts the onset from $N_{\rm part} \simeq 320$ to about 190,
as also shown in Fig.\ \ref{dir-psi}.

\begin{figure}[htb]
\hskip -0.5cm
\centerline{\psfig{file=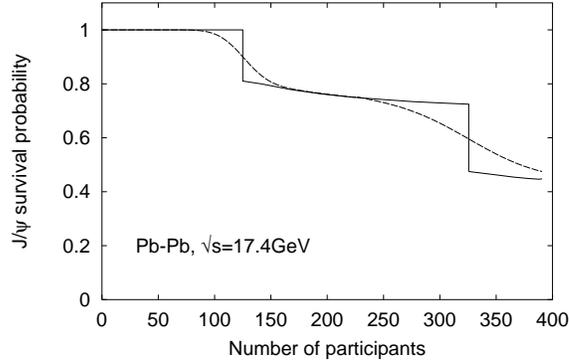,width=7.5cm}}
\caption{\J~ survival pattern in Pb-Pb collisions at the SPS, as function 
of centrality.}
\label{survival-Pb}
\end{figure}

We now combine both thresholds and use the nuclear collision profile to 
determine the distribution of charmonium production in the transverse 
plane \cite{KLNS}, in order to determine which are formed in the cluster 
and which are not. As function of centrality, this leads to the survival 
pattern shown in Fig.\ \ref{survival-Pb}. 

It is clear that various theoretical and
experimental sources (fluctuations, resolution, binning, etc.) will lead 
to some smearing of the theoretical survival pattern. To obtain
some idea of what this may lead to, we have folded our calculated
distribution $S(N_{\rm part})$ with a Gaussian smearing function, 
\be
S(N_{\rm part}) = \int d{\bar N}_{\rm part}~ S({\bar N}_{\rm part})
{1 \over \sqrt{2\pi}~ \sigma} 
\exp\{-(N_{\rm part}-{\bar N}_{\rm part})^2 / 2\sigma^2\},
\label{5}
\ee
using a relative resolution of $\sigma/N_{\rm part}$ = 15 \% at
the respective thresholds. The result is also shown in 
Fig.\ \ref{survival-Pb}.

\medskip

\begin{figure}[htb]
\mbox{
\epsfig{file=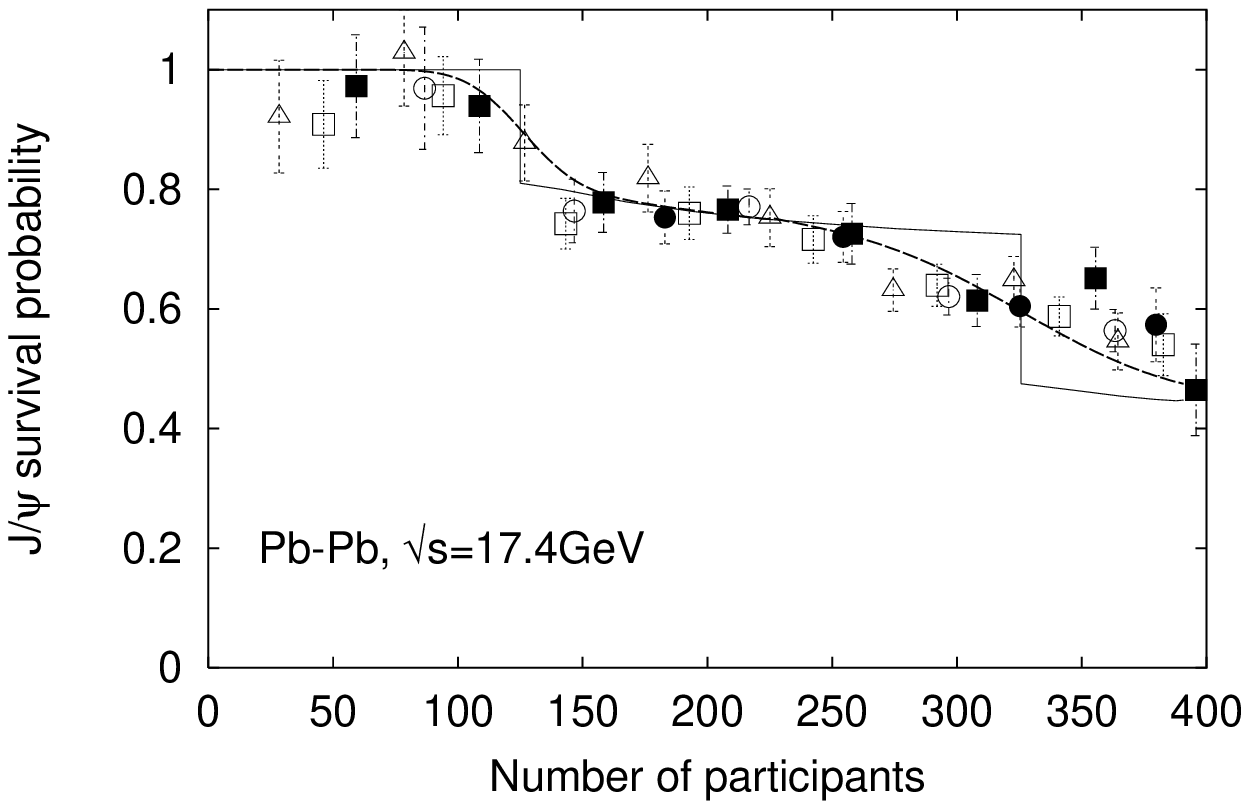, width=7.5cm}
\hskip 0.5cm
\epsfig{file=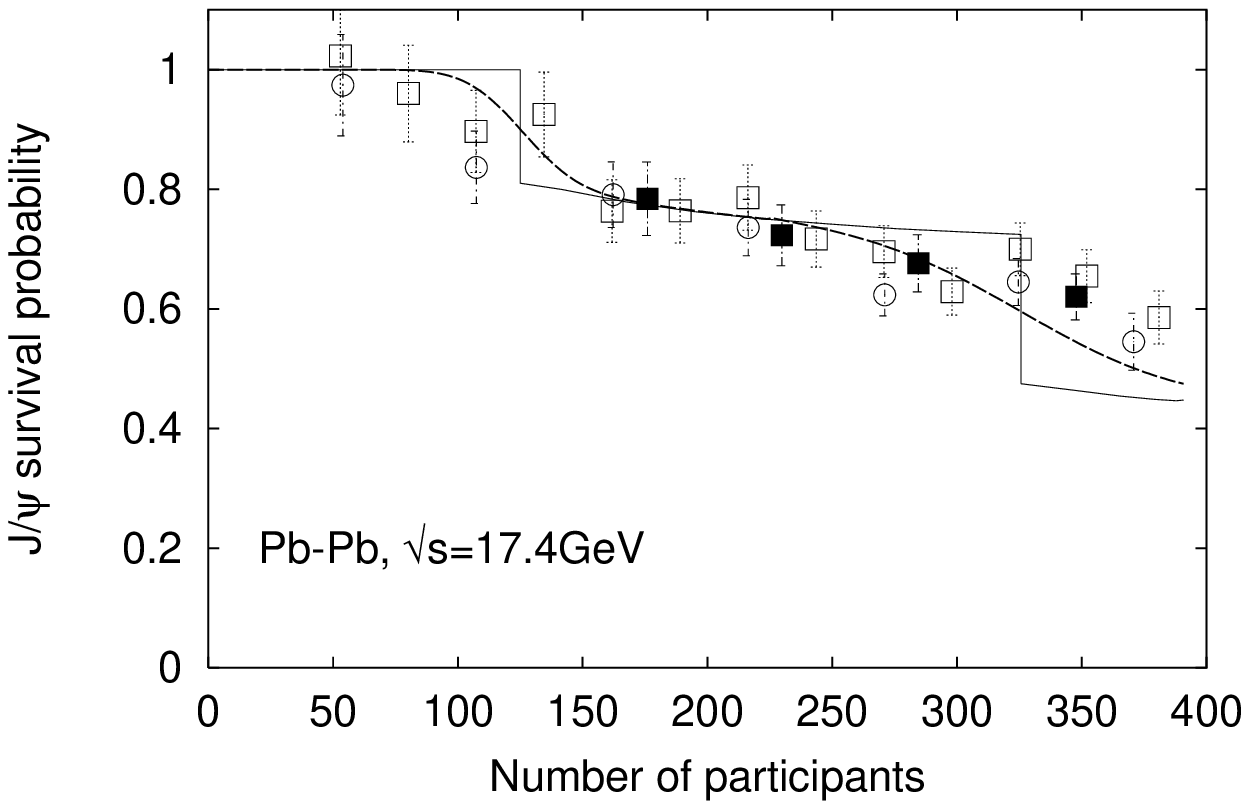, width=7.5cm}}
\caption{\J~ survival pattern in Pb-Pb collisions at the SPS, as function 
of centrality, determined from $E_T$ (left) and from $E_{ZDC}$ (right).}
\label{survival-Pb-data}
\end{figure}

The centrality dependence of the \J~survival has been measured
in Pb-Pb collisions at the SPS using two independent variables, the
associated transverse energy, $E_T$, and the energy of the undeflected
nucleons, $E_{\rm ZDC}$. 
The first effectively measures the number of hadronic 
secondaries in the transverse plane, the second
the number of participating nucleons. We want to study anomalous
charmonium suppression, i.e., effects above and beyond the normal 
suppression already observed in p-A collisions, due to absorption of
the nascent charmonium state in normal nuclear matter \cite{K-S2}.
To determine this normal suppression, we use a recent analysis 
\cite{Cortese} of $\!p-\!A$ and $S\!-\!U$
data by the NA50 group, which finds a somewhat
weaker nuclear absorption, $\sigma_d \simeq 4.3 \pm 0.6$, than previous
studies. Normalizing the Pb-Pb data to the normal nuclear suppression
thus specified leads to the points shown in Fig. \ref{survival-Pb-data},
where they are compared to our parton percolation suppression pattern.  
We see good agreement between the calculated curve and the data points. 
We recall, however, that the second threshold value depends crucially on
the \J~radius. 

\medskip

\medskip

\begin{figure}[htb]
\mbox{
\epsfig{file=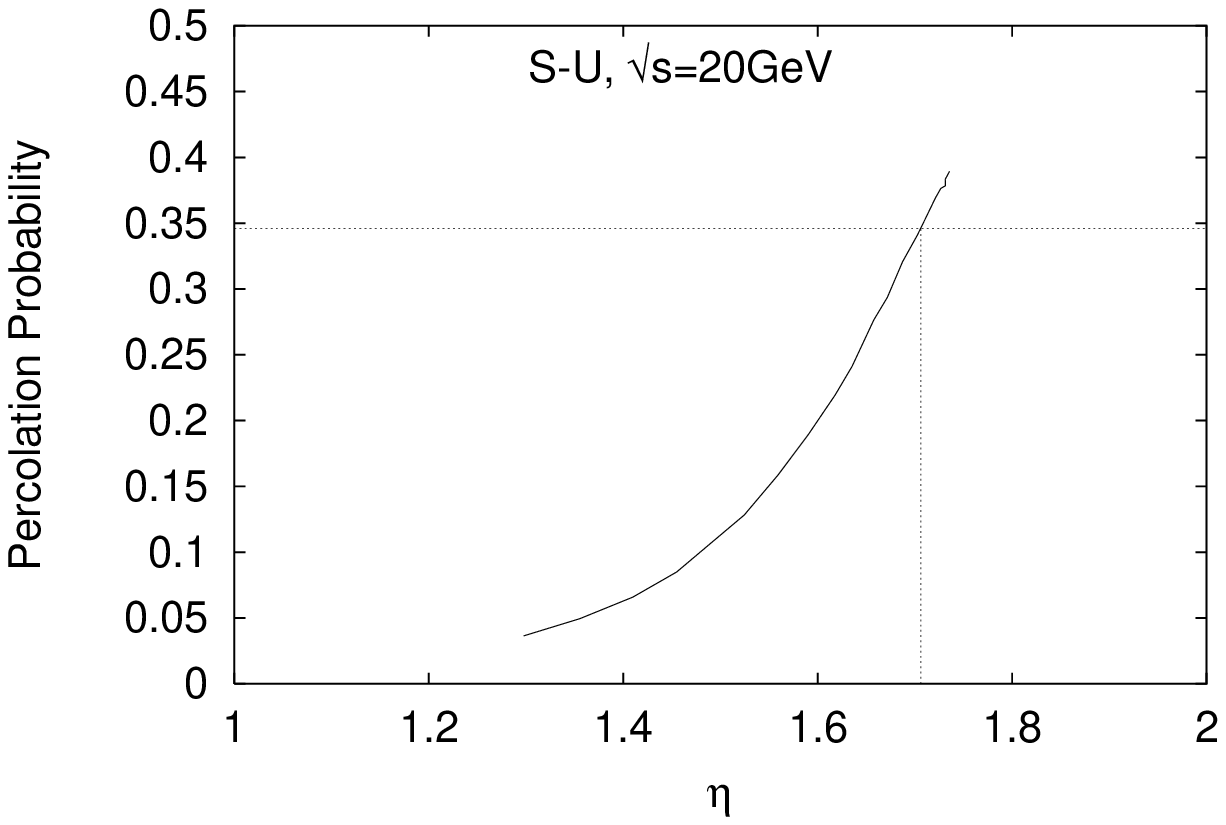,width=7.5cm}
\hskip0.5cm
\epsfig{file=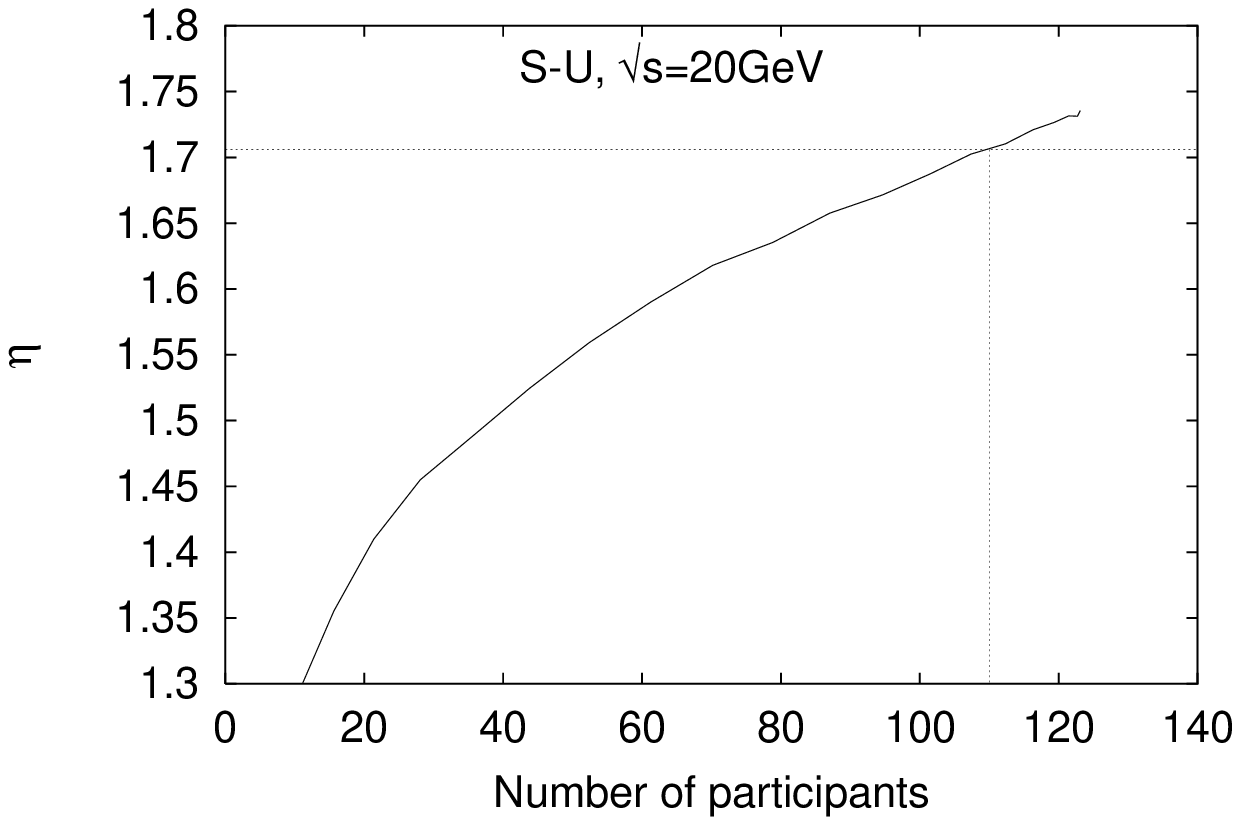,width=7.5cm}}
\vskip-0.2cm
\caption{Percolation onset in S-U collisions at the SPS.}
\label{S-chi}

\vskip 1cm

\centerline{\psfig{file=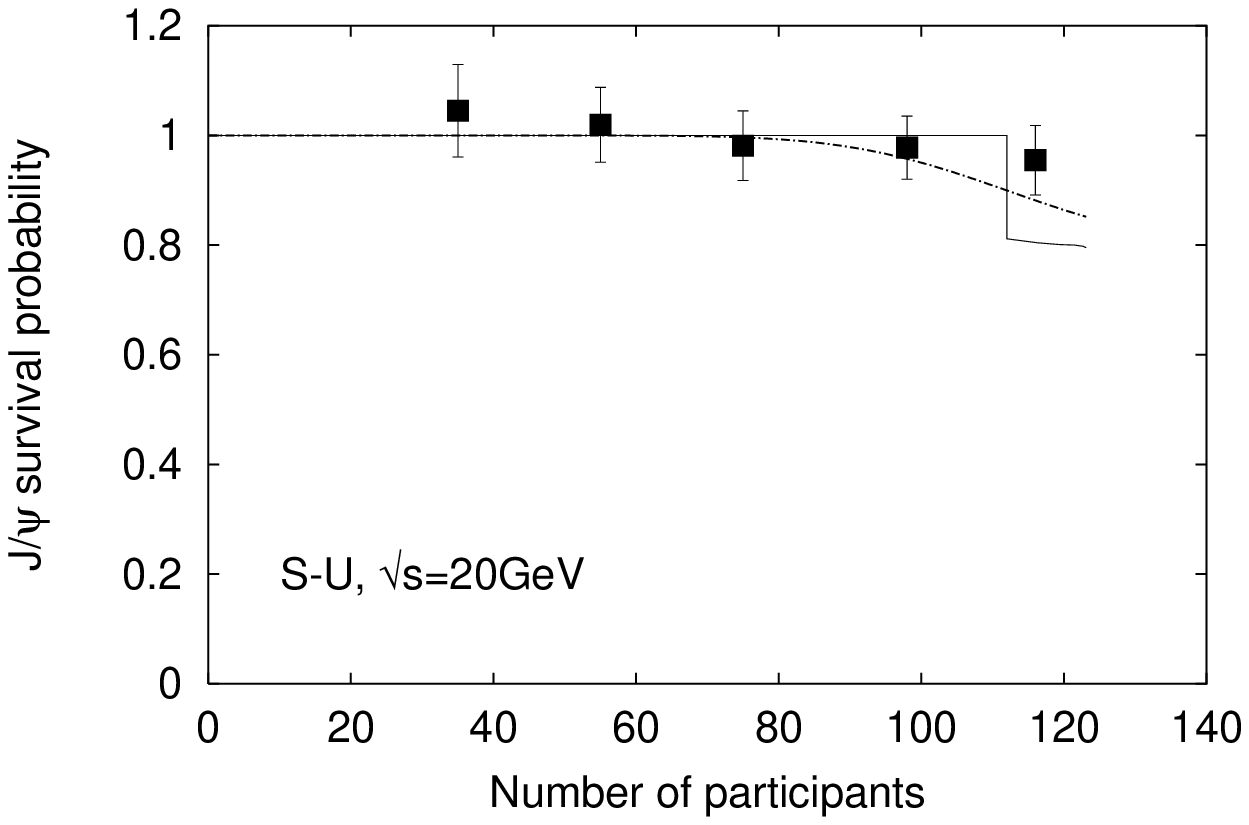, width=7.5cm}}
\vskip-0.2cm
\caption{\J~ survival pattern in S-U collisions at the SPS, as function 
of centrality.}
\label{survival-S}
\end{figure}

Next we want to check the behaviour predicted for $S\!-\!U$ collisions
at the SPS, since this shows at best only very slight anomalous
suppression\footnote{Initially, it was claimed that the data
are fully accounted for by normal nuclear absorption. However,
the lower 
absorption cross sections obtained in the mentioned new p-A studies
allow a slight anomalous suppression for the most central $S\!-\!U$ 
collisions.}. In Fig.\ \ref{S-chi}, we show the percolation behaviour, 
and in Fig.\ \ref{survival-S}, the resulting survival pattern together 
with the
data \cite{S-U-data}. The threshold obtained here corresponds to the 
general onset of 
percolation; the resolution scale of the direct \J~cannot be reached 
in $S\!-\!U$ collisions. The corresponding smeared form of the calculated 
distribution is also included.

\medskip

We close this section with some general comments on the suppression
of charmonium production. We have here considered parton percolation 
as mechanism for anomalous \J~suppression, defined as the reduction of 
the observed production rate beyond that predicted by normal absorption 
in standard nuclear matter. In doing so, we neglect other conceivable
suppression mechanisms, in particular dissociation by (non-percolating)
secondary hadronic
or partonic comovers. This is not expected to play a significant role
for the suppression of $1S$ states, i.e., directly produced \J's:
the large binding energy (650 MeV) makes a dissociation by
confined partons very unlikely \cite{K-S1}. However,
since about 40 \% of the observed \J's come from the decay of higher
excited states (feed-down from \X~and \P~production), we have to
consider the fate of these as well. The \X~($\sim$ 250 MeV) is still 
rather strongly bound, but the much smaller \P~binding energy 
($\sim$ 60 MeV) means that any interaction will cause break-up. 
It is thus not surprising that the \P~suffer an anomalous suppression 
even in the most peripheral S-U interactions \cite{Psi'-SU}.
This is most likely not related to any parton deconfinement and 
has to be addressed separately. 

\bigskip

\noindent{\bf \large 3.\ Predictions for NA60 and RHIC}

\bigskip

\begin{figure}[htb]
\mbox{
\epsfig{file=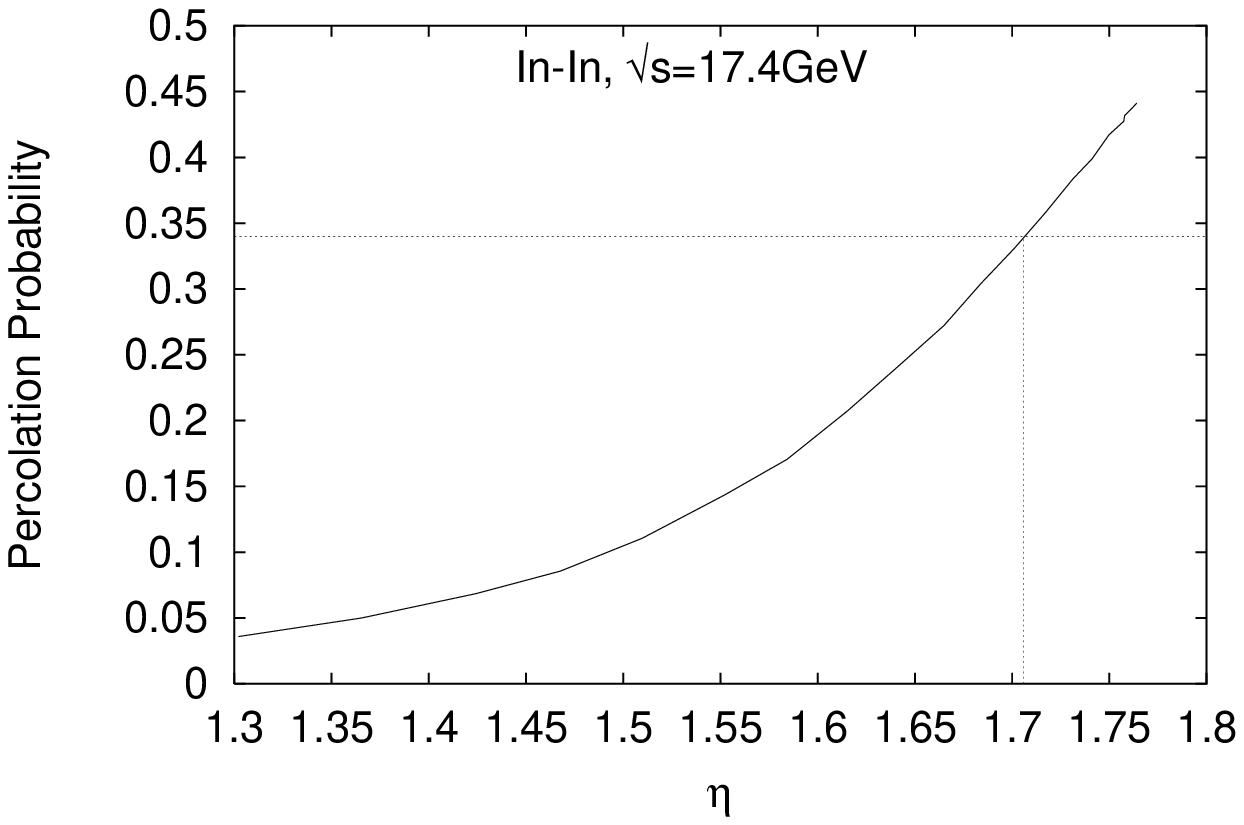,width=7.5cm}
\hskip0.5cm
\epsfig{file=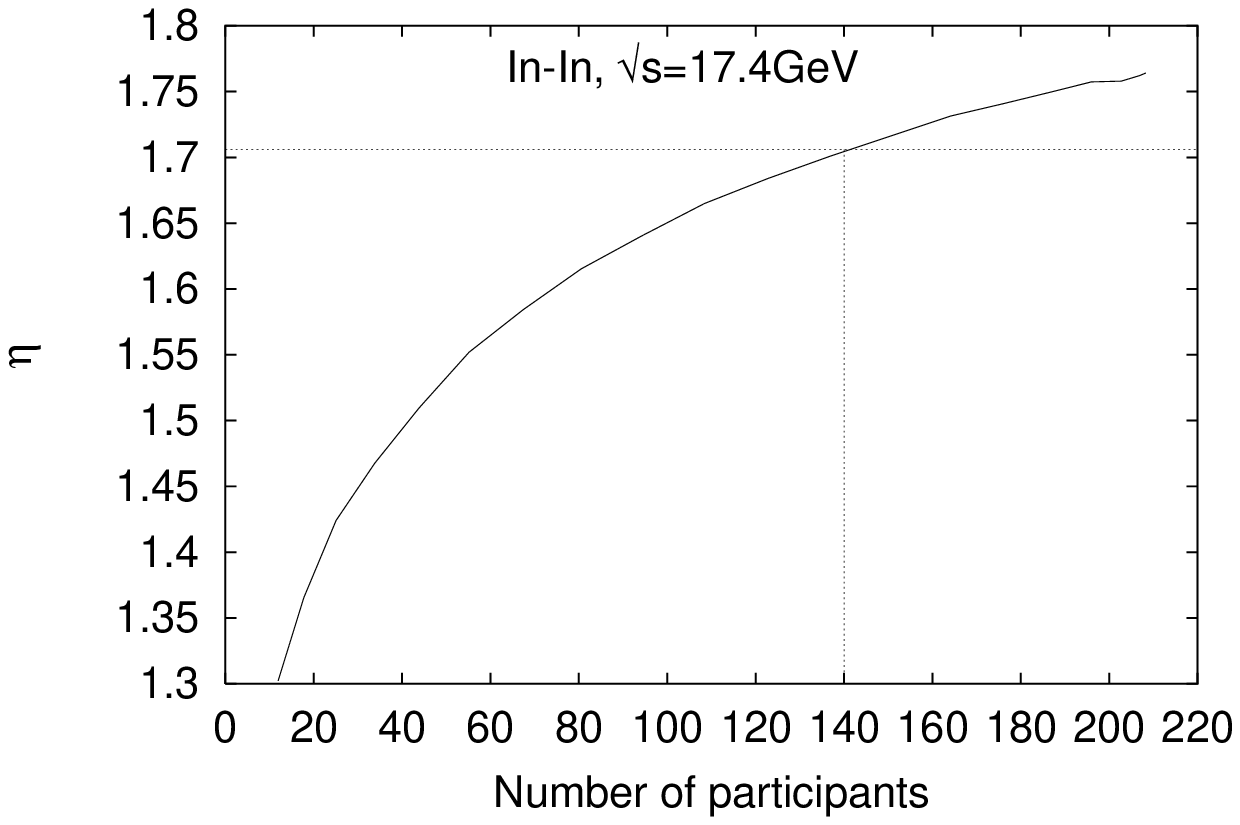,width=7.5cm}}
\caption{Percolation onset in In-In collisions at the SPS.}
\label{In-chi}

\vspace*{1cm}

\centerline{\psfig{file=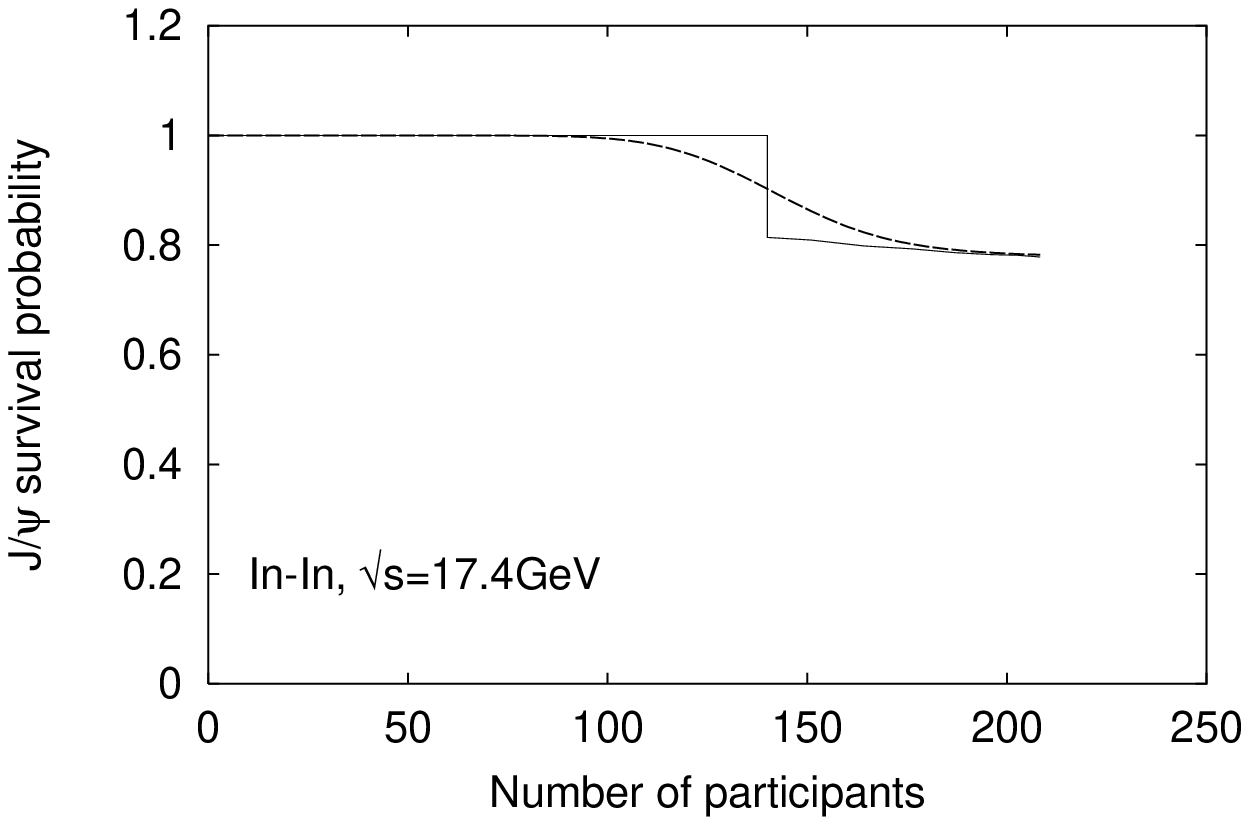,width=7.5cm}}
\caption{\J~ survival pattern in In-In collisions at the SPS, as function 
of centrality.}
\label{survival-In}
\end{figure}

We now turn to the forthcoming $In\!-\!In$ collisions at the SPS.
The percolation pattern is shown in Fig.\ \ref{In-chi}; the general 
percolation threshold is reached at $N_{\rm part} \simeq 140$. In
contrast to $Pb\!-\!Pb$ collisions,
the threshold for directly produced \J's here remains above
the produced density even for the most central collisions
(but recall the caveat concerning its dependence on the \J~radius). The
resulting survival pattern for  $In\!-\!In$ collisions at the SPS 
is given in Fig.\ \ref{survival-In}, including also the smeared form.
We see that the most central 30 \% of the collisions show a 
suppression of the \J's produced by feed-down from higher excited 
charmonium states.


\medskip

\begin{figure}[htb]
\mbox{
\epsfig{file=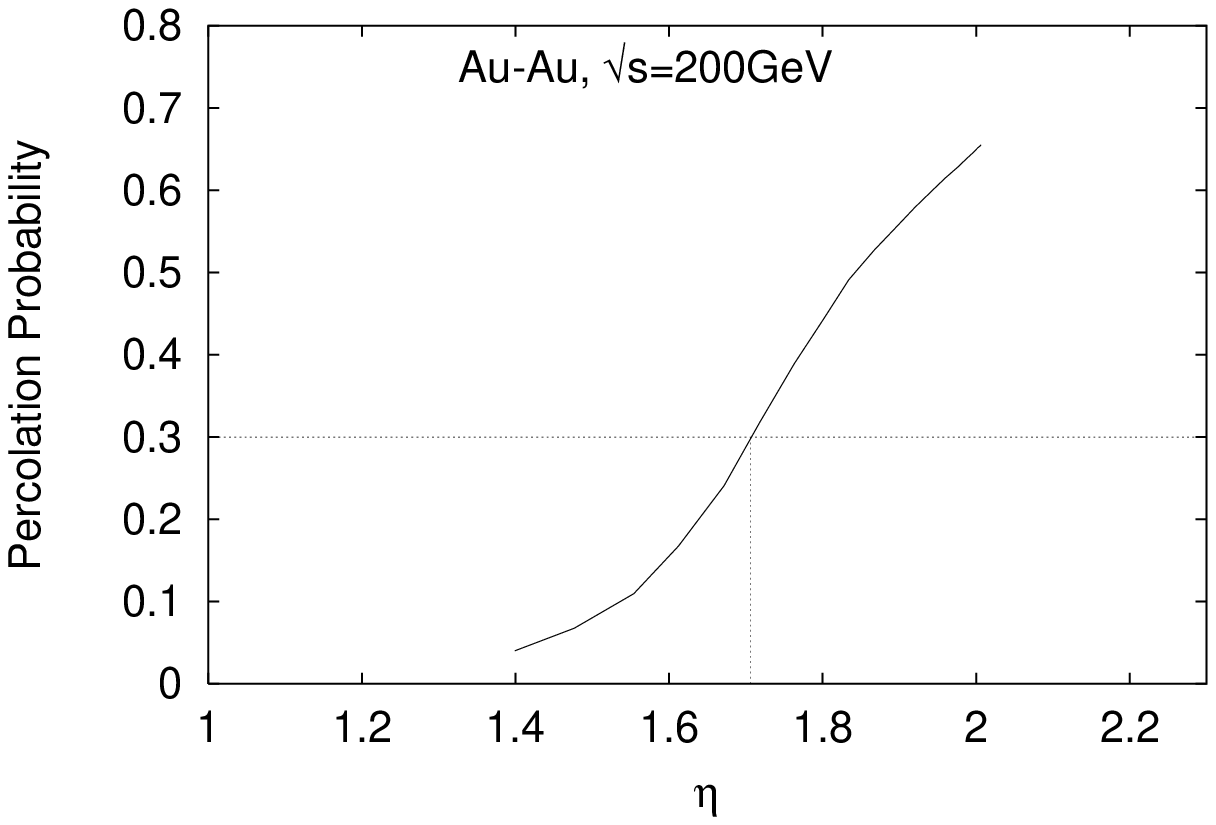,width=7.5cm}
\hskip0.5cm
\epsfig{file=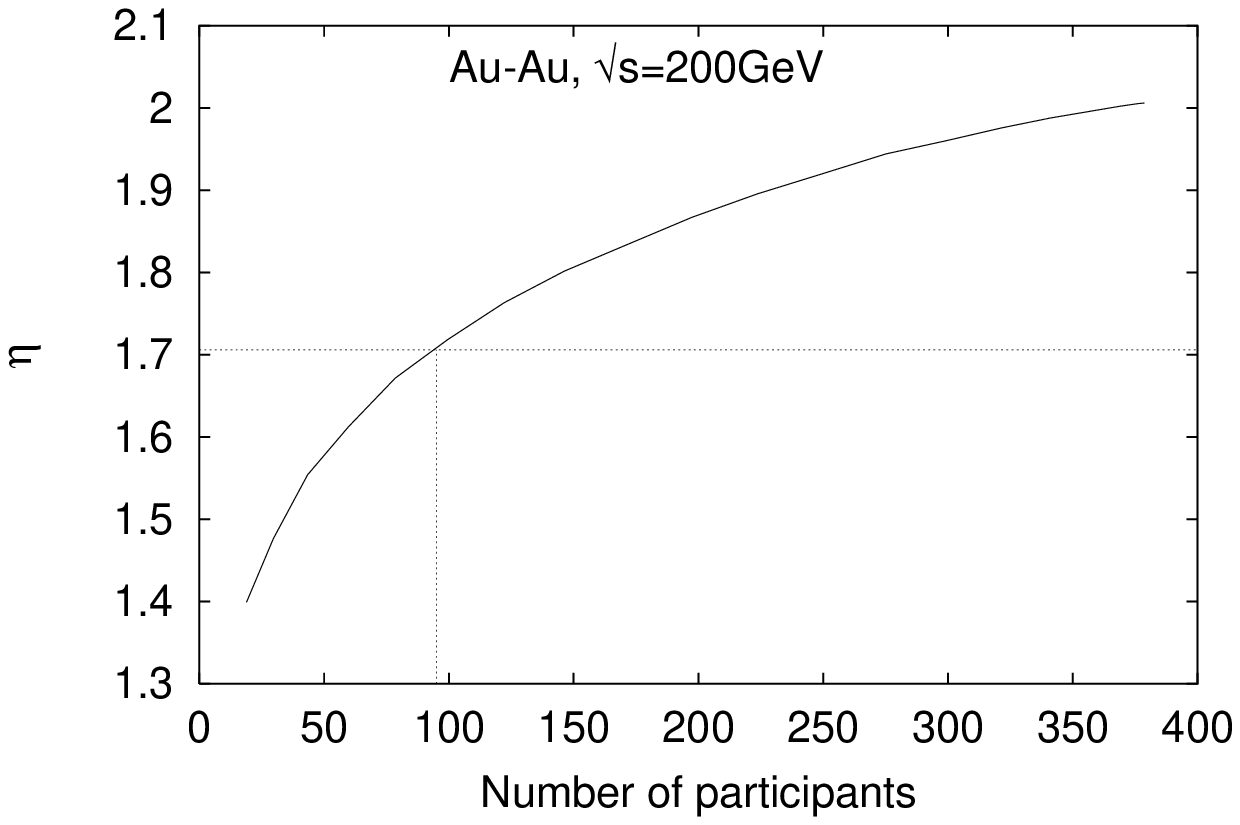,width=7.5cm}}
\vskip -0.2cm
\caption{Percolation onset in Au-Au collisions at RHIC.}
\label{Au-all}

\vspace*{1cm}

\hskip -0.5cm
\centerline{\psfig{file=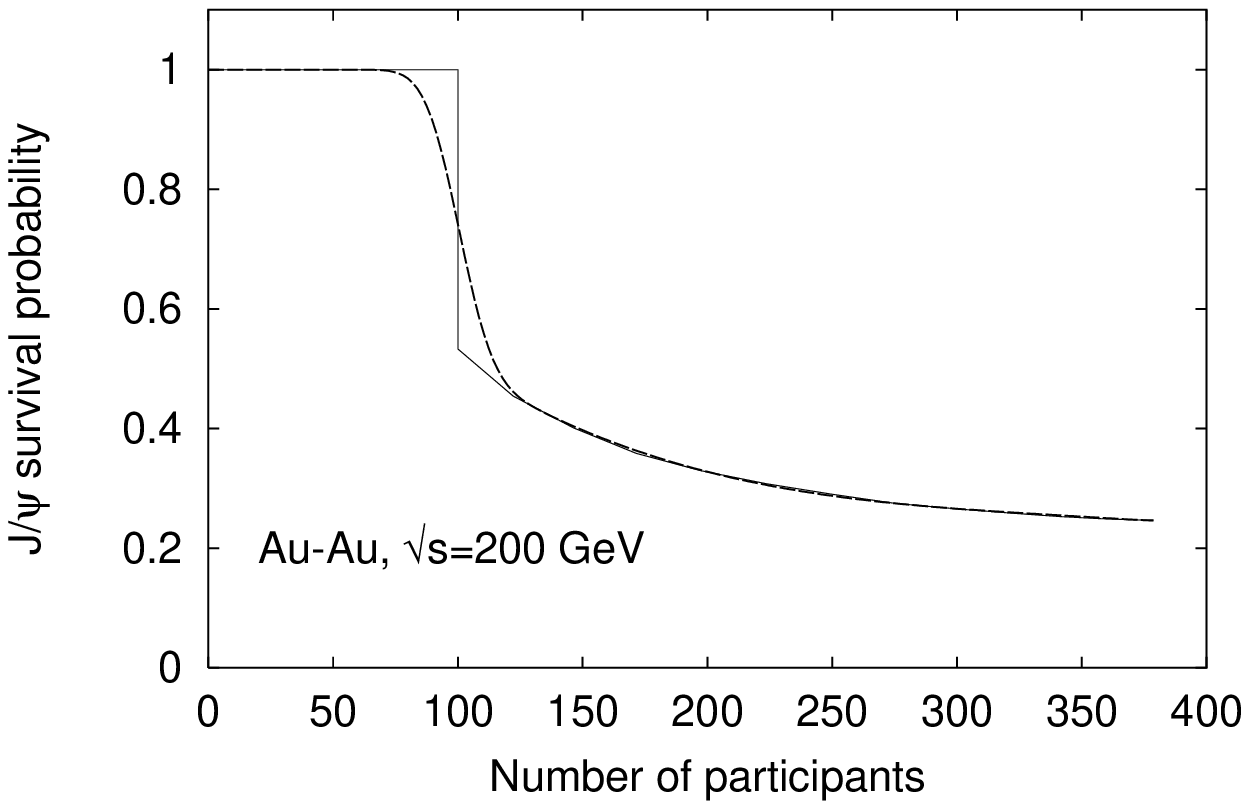,width=7.5cm}}
\vskip -0.2cm
\caption{\J~ survival pattern in Au-Au collisions at RHIC, as function 
of centrality.}
\label{survival-Au}
\end{figure}

Finally we address the suppression pattern for $Au\!-\!Au$ collisions
at RHIC, with the percolation pattern shown in Fig.\ \ref{Au-all}.
Here the increased parton density shifts the onset of percolation to
a higher resolution scale, so that from the threshold on, all charmonium
states are suppressed. This leads to a single step suppression pattern,
starting at $N_{\rm part} \simeq 90$; it is given in Fig.\ 
\ref{survival-Au}. Recall that we are here specifying the predicted
anomalous suppression, beyond the normal suppression resulting from
(pre-resonance) absorption in standard nuclear matter. For various reasons,
this normal suppression may depend on the collision energy
and can thus be quite different from that observed at SPS energy. 
To determine the anomalous suppression at RHIC, it is therefore crucial 
to first measure the normal absorption in $p\!-\!A$ (or $d\!-\!A$) 
collisions. The threshold for $Au\!-\!Au$ collisions as obtained here
occurs for considerably more central collisions than found
in previous studies \cite{torino}, in which the dependence of parton 
distribution functions on the resolution scale had not taken into account.

\bigskip

In summary: if the origin of the two-step suppression pattern
observed in the $Pb\!-\!Pb$ 
SPS collisions is indeed the onset of parton percolation at different
resolution scales, as discussed, then we expect a single step pattern
both in  $In\!-\!In$ collisions at the SPS and in $Au\!-\!Au$ collisions
at RHIC. In the former, the step is due to the suppression of the 
higher excited states only, with directly produced \J's surviving;
in the latter, there is a common threshold for all charmonium states
and hence only one step.

\bigskip

\noindent{\bf \large 4.\ Discussion}

\bigskip 

We have presented the \J~suppression pattern as obtained from parton
percolation under the assumption of complete charmonium dissociation 
within a percolating cluster of sufficient resolution scale, full survival 
in media below percolation density or of insufficient resolution scale.
In this closing section, we want to summarize the uncertainties
of such an approach and compare its outcome to that of a thermal description. 

\medskip 

The onset of parton percolation is specified directly through the
parton distribution functions (PDF's) as obtained from an analysis of deep 
inelastic lepton-hadron data. Uncertainties in the PDF's produce some
uncertainty in the precise values of the percolation threshold, but
not in the onset as such. As mentioned, percolation must occur in
any approach based on partons with an intrinsic transverse momentum. 
The assumption tested by comparing our results to \J~data is that for 
charmonium suppression in nuclear collisions, the primary parton density 
is the crucial variable and the formation of a percolating cluster the 
relevant mechanism. 

\medskip

As noted, the onset of the suppression of directly produced \J's
at SPS energy is quite sensitive to the choice of charmonium parameters
(and on that of the parton distribution functions, for which slight
changes would also produce considerable threshold shifts). On the other
hand, the onset of the first step in SPS $Pb\!-\!Pb$ pattern, the predicted 
single step in SPS $In\!-\!In$ collisions, and the predicted single step 
in RHIC $Au\!-\!Au$ collisions should be quite robust. This should
help in identifying the origin of the second step in SPS $Pb\!-\!Pb$
collisions. It has been suggested \cite{B-O} that this drop is largely 
due to transverse energy fluctuations for the most central collisions;
this would imply that it should be absent in zero-degree calorimeter
studies, where (see Fig.\ \ref{survival-Pb-data}) it seems to still occur 
to some extent \cite{E-ZDC}. A 
further test should come from the SPS $In\!-\!In$ and the RHIC 
$Au\!-\!Au$ data, where the existence of a second step could not be 
attributed to parton percolation.

\medskip

Finally we turn briefly to the question of the distinguishing the initial 
state charmonium suppression presented here from a suppression occurring 
at a later thermalized stage. We had here assumed that
the primary parton configuration determines the pattern of \J~suppression. 
What would one expect if the initial parton configuration leads to
subsequent thermalization, with the formation of a more or less
equilibrated quark-gluon plasma, and it is this QGP which causes
the observed \J~suppression? The most recent (quenched) lattice QCD studies 
\cite{Biele,Japan} indicate that in a hot thermal medium of quarks and 
gluons, higher excited charmonium states are dissociated at or below $T_c$. 
In contrast, the ground state \J~survives to considerably higher 
temperatures; present calculations still show a clear \J~signal 
at least up to $T=1.5~T_c$. This would imply that thermal dissociation 
removes the feed-down contributions from \P~and \X~states
for energy densities around or below 1 GeV/fm$^3$, while
energy densities of more than 6--8 GeV/fm$^3$ are
required for the melting of directly produced \J's.
It seems difficult to accommodate the existing SPS data 
with these thresholds, while the pattern resulting from parton
percolation can do so quite naturally. For $Au\!-\!Au$ collisions at 
RHIC, thermal melting would suppress the higher state 
feed-down contributions for practically all centralities, 
while the directly produced \J's disappear at best for the most
central collisions. We thus conclude that the charmonium suppression 
thresholds from initial state and from thermal suppression 
are quite different also at much higher energy.

\vskip 0.8cm

\centerline{\large \bf Acknowledgements}

\bigskip

It is a pleasure to thank F.\ Karsch, E.\ Laermann, C.\ Louren{\c c}o
and P.\ Petreczky for stimulating remarks and discussions.

\end{document}